\documentclass[aps,prl,twocolumn,showpacs,floatfix]{revtex4-1}

\usepackage{amssymb,amsmath}
\usepackage{graphicx}
\usepackage{eurosym}
\usepackage{color}
\usepackage[normalem]{ulem}
\usepackage[makeroom]{cancel}

\newcommand{\ba}{{\bf a}}
\newcommand{\be}{{\bf e}}
\newcommand{\br}{{\bf r}}

\newcommand{\bA}{{\bf A}}
\newcommand{\bE}{{\bf E}}

\newcommand{\cH}{\mathcal{H}}
\newcommand{\cQ}{\mathcal{Q}}

\newcommand{\tb}{\tilde{b}}

\newcommand{\tbA}{\tilde{\bf A}}

\newcommand{\p}{{\cal P}}
\newcommand{\PT}{{\cal PT}}
\newcommand{\T}{{\cal T}}
\newcommand{\C}{{\cal C}}
\newcommand{\K}{{\cal K}}

\newcommand{\cE}{{ { \cal E}}}

\begin{document} 

\title{Odd-time reversal $\PT$ symmetry induced by anti-$\PT$-symmetric medium}

 \author{Vladimir V. Konotop$^1$ and Dmitry A. Zezyulin$^{2}$}
 \affiliation{$^1$Departamento de F\'isica and Centro de F\'isica Te\'orica e Computacional, Faculdade de Ci\^encias, Universidade de Lisboa, Campo Grande 2, Edif\'icio C8, Lisboa 1749-016, Portugal
 	\\
 	$^2$ITMO University, St.~Petersburg 197101, Russia 
 }

\begin{abstract} 
We introduce an optical system (a coupler) obeying parity-time ($\PT$) symmetry with {\em odd}-time reversal, $\T^2=-1$. It is implemented with two birefringent waveguides embedded in an anti-$\PT$-symmetric medium. The system possesses properties, which are untypical for most physical systems with the conventional even-time reversal. Having symmetry-protected degeneracy of the linear modes, the coupler allows for realization of a coherent switch operating with a superposition of binary states  which are distinguished by their polarizations. When a Kerr nonlinearity is taken into account, each linear state, being double degenerated,  bifurcates into several distinct nonlinear modes, some of which are dynamically stable. The nonlinear modes are characterized by   amplitude and by polarization and come in $\PT$-conjugate pairs.  
\end{abstract}
\maketitle

\paragraph{Introduction.} 
The concepts of parity ($\p$) and time ($\T$) symmetries, 
intensively discussed in the context of 
non-Hermitian quantum mechanics since the seminal work~\cite{BenderBoet},  nowadays acquired great significance in practically all areas of physics dealing with 
linear and nonlinear wave phenomena~\cite{review}. Universality of the paradigm, first recognized in optics~\cite{Muga,disc_opt,Christodoulides}, is based on the mathematical similarity between the parabolic equation describing light propagation in various settings and  the Schr\"odinger equation governing dynamics of a non-relativistic quantum particle.  Respectively, the parity and time inversion operators used in   most of the applications had the same form as those 
for a spinless  quantum particle, i.e., $\p\psi(\br,t)=\psi(-\br,t)$ and $\T\psi(\br,t)=\psi^*(\br,-t)$. From the theoretical point of view, however, the operators $\p$ and $\T$ can have much more general form 
\cite{BendManh}. As a matter of fact,  
various definitions of the parity operator, which is an involution, i.e. satisfies $\p^2=1$, 
have  already been explored in discrete optics. For instance,  for dimer models the operator  $\p$ is tantamount
to the $\sigma_1$ Pauli matrix~\cite{disc_opt}, and in more complex quadrimer and oligomer models $\p$ can be defined as Kronecker products of Pauli matrices~\cite{general_p,ZK,review}. The time reversal operator $\T$ 
is anti-linear and, in quantum mechanics, it is  even for bosons, $\T^2=1$,  and odd for fermions, $\T^2=- 1$ \cite{Messiah}. However, only the former possibility was used in all classical applications (i.e., beyond quantum mechanics) of the non-Hermitian physics.  

The non-Hermitian quantum mechanics with odd time reversal, $\T^2=-1$, has been brought to the discussion by a series of works initiated by~\cite{SmithMathur,BendKlev}.  
The respective Hamiltonians obey interesting properties (some of them are recalled below) which, however, 
have never been  explored in other physical applications. This leads to the first goal of this Letter, which is to  introduce an optical system obeying {\em odd} $\PT$ symmetry. We illustrate the utility of   
such a system   with two examples. First, we propose a coherent optical switch which operates with linear superpositions of binary states, rather than with single states, as the conventional switches based on even $\PT$-symmetry do~\cite{switch}. Second, we describe peculiarities of nonlinear modes in odd-$\PT$ systems, where the nonlinearity is odd-$\PT$ symmetric, too.

Let us also recall other recent developments in optics of media with special symmetries.  It was suggested in~\cite{Ge} to explore properties of  anti-$\PT$-symmetric optical media which are characterized by dielectric permittivities with  $\varepsilon(\br)=-\varepsilon^*(-\br)$ and can be realized, say, in metamaterials. More recently, experimental realization  of anti-$\PT$-symmetric media in atomic vapors has been reported in~\cite{Peng2016},  and other schemes implementing the idea with dissipatively coupled optical systems, have been designed in \cite{anti-PT}. Practical applications of such media, however, remain unexplored. Thus, the second  goal of this Letter is to show that an anti-$\PT$ symmetric medium is a natural physical environment where  the odd $\PT$ symmetry can be realized.   

\begin{figure}
	\centering
		\includegraphics[width=1.0\columnwidth]{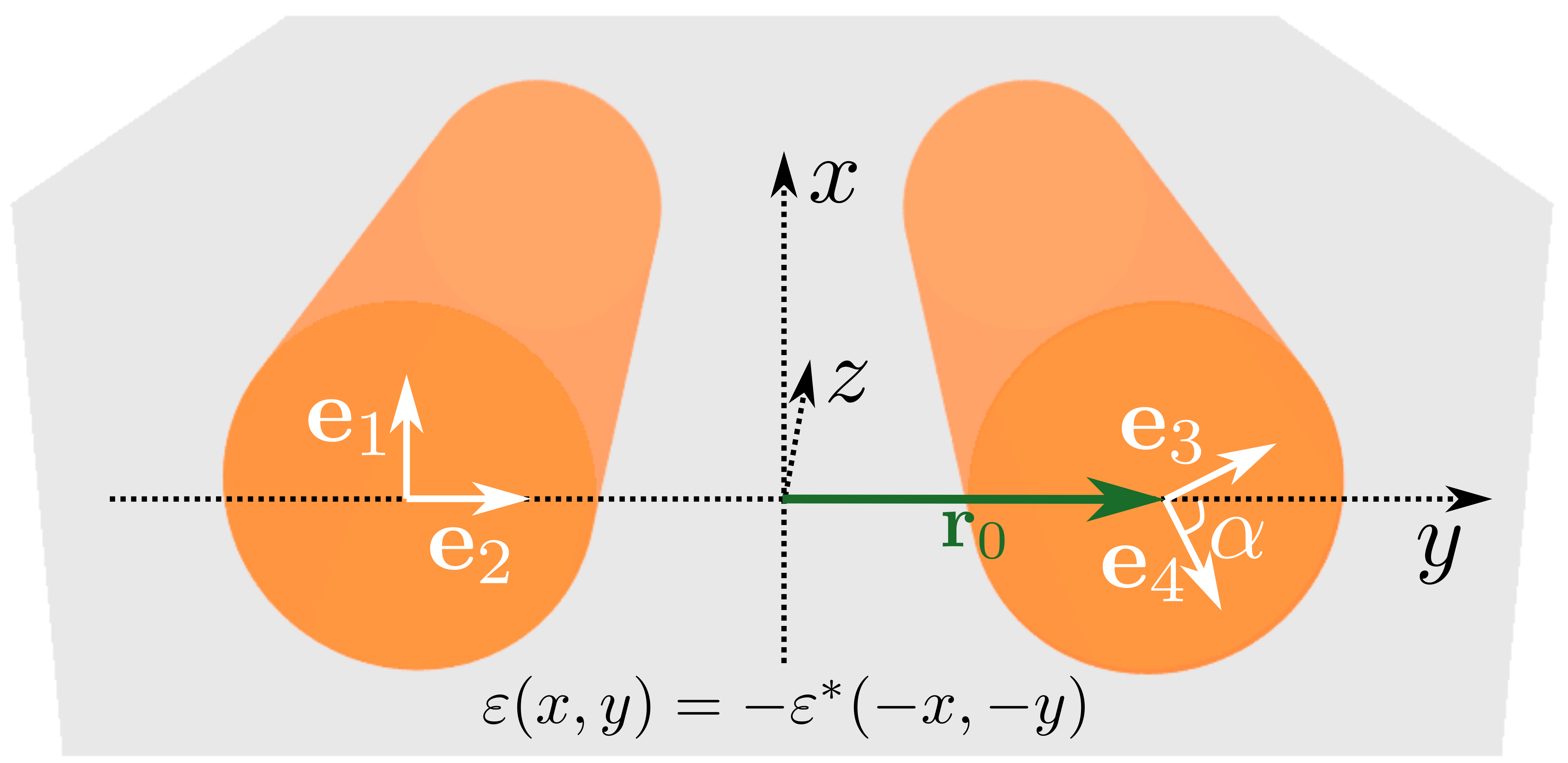}
	\caption{
		Coupled transparent waveguides embedded in an anti-$\PT$-symmetric medium   (see the text for notations).}
	\label{fig:one}
\end{figure}

\paragraph{Optical coupler with odd $\PT$ symmetry.} 
Consider a system of two birefringent waveguides, each one with orthogonal principal axes. To simplify the model, we neglect a mismatch between propagation constants of the polarizations inside each waveguide, but take into  account   a mismatch $2\delta$ between the propagation constants of the waveguides: $q_{1,2}= q \mp\delta$, where $q$ is the average propagation constant.  Let these waveguides be coupled to each other by an isotropic medium with active and absorbing domains as schematically shown in Fig.~\ref{fig:one}.
The components of the guided monochromatic electric fields  can be written as
$
 \bE_{1} =[\be_1 A_1(z) \psi_{1}(\br)+\be_2 A_2(z) \psi_{2}(\br)]e^{i(q-\delta)z}
$ 
and $
\bE_{2} =[\be_3A_3(z)\psi_{3}(\br)+\be_4A_4(z)\psi_{4}(\br)]e^{i(q+\delta)z}
$,
where $\br=(x,y)$, $\be_{j}$ and $\psi_{j}(\br)$ are the polarization vectors and the respective transverse distributions of the modes, $A_j(z)$ are slowly varying field amplitudes which depend on the propagation distance $z$. Polarization axes in each waveguide are orthogonal, $\be_1\be_2=\be_3\be_4=0$, and in different waveguides are mutually rotated by angle $\alpha$, ensuring the relations $\be_1\be_3=\be_2\be_4=\cos\alpha $ and $\be_1\be_4=-\be_2\be_3=-\sin\alpha$. The modes are weakly guided, so that the same polarization properties hold for the fields outside the waveguides  cores.

Since $\be_1$ is orthogonal to $\textbf{e}_2$ and $\be_3$ is orthogonal to $\textbf{e}_4$, the coupling is possible only between one polarization in a given waveguide and  two polarizations in another one. Such a coupling is determined by the overlapping integrals   
$
\kappa_{jk}=\be_j\be_k\int \psi_{j}^*(\br)\varepsilon(\br)\psi_{k}(\br)d^2\br,
$ 
 where $j,k=1,...,4$.

Let the medium in which the waveguides are embedded be anti-$\PT$-symmetric:  $\varepsilon(\br)=-\varepsilon^*(-\br)$. Assuming that $\psi_{j}(\br)\approx \psi_{j}(|\br|)$, i.e., the transverse field distribution is approximately radial, one ensures that $\kappa_{jk}=-\kappa_{kj}^*$ where $j=1,2$ and $k=3,4$. To further simplify the model, we consider the transverse distributions to differ only by phase mismatches $\varphi$ and $\vartheta$, according to the relations $\psi_{1}(\br)e^{i(\varphi+\vartheta)/2}=\psi_{2}(\br)e^{i(\vartheta-\varphi)/2}\equiv\psi(\br+\br_0)$ and $\psi_{3}(\br)e^{i(\varphi+\vartheta)/2}=\psi_{4}(\br)e^{i(\vartheta-\varphi)/2}\equiv\psi(\br-\br_0)$, where $\pm \br_0$ are the   coordinates of the core centers (see Fig.~\ref{fig:one}). Thus, for the coupling coefficients we have
$\kappa_{13}=\kappa_{24}=i \kappa   \cos\alpha$ and  $\kappa_{14}=\kappa_{23}^*=-i\kappa e^{i\varphi} \sin\alpha$, 
where
$
\kappa =-i \int \psi^*(\br-\br_0)\varepsilon(\br)\psi(\br+\br_0)d^2\br
$
is real.
If the waveguides possess Kerr nonlinearity, one can write the system describing the evolution of the slowly varying amplitudes $\bA=
\left(
A_1,  A_2,  A_3, A_4
\right)^T$ ($T$ stands for transpose) in the matrix from  \cite{supplement}
\begin{eqnarray}
\label{main_dynamic}
i\dot{\bA}=H_\delta \bA -F(\bA)\bA,  \quad H_\delta=\left(\begin{array}{cc}
\delta \sigma_0 & i \kappa C
\\
i \kappa C^\dagger & -\delta \sigma_0
\end{array}\right).
\end{eqnarray} 
Here   $\sigma_0$ is the  $2\times 2$ identity matrix,  $C$ is the coupling matrix
\begin{eqnarray} 
\label{Hamilt} 
 C=\left(\begin{array}{cc}
  e^{-i\vartheta}\cos \alpha & -e^{i\varphi} \sin \alpha
\\
e^{-i\varphi} \sin\alpha & e^{i\vartheta}\cos\alpha 
\end{array}\right),
\end{eqnarray} 
and the nonlinearity has the form known for birefringent waveguides~\cite{Menyuk}:
\begin{eqnarray}
\label{nonlin}
F(\bA)=\mbox{diag}\left(|A_1|^2+\frac{2}{3}|A_2|^2,|A_2|^2
 +\frac{2}{3}|A_1|^2, 
 \right. \nonumber  \hspace{0.5cm}\\ \left.  
 |A_3|^2+\frac{2}{3}|A_4|^2,|A_4|^2+\frac{2}{3}|A_3|^2 \right).
\end{eqnarray}

The main feature of coupler (\ref{main_dynamic}), explored below, is that  the coupling matrix $C$ is a real quaternion~\cite{supplement}. Recalling the known results~\cite{SmithMathur,BendKlev}, one concludes that $H_\delta$ obeys odd $\PT$ symmetry with parity operator $\p=
\gamma^0$, where $\gamma^0$ is the Dirac gamma matrix, and time reversal $\T=\sigma_0\otimes (i\sigma_2)\K$, where $\K$ is the element-wise complex conjugation (note that $i\sigma_2\K$ is the usual time reversal operator for spin-1/2 fermions~\cite{Messiah}). The relevant properties of the introduced operators are $\p^2 = 1$, $\T^2 = -1$,   $[\p,\T]=0$, and 
$[\PT,H_\delta]=0$.

We start the analysis of system (\ref{main_dynamic}) with the linear limit, $F(\bA)\equiv 0$. The guided modes are described by the eigenvalue problem: $\tb \tbA=H_\delta\tbA$ (we use tildes for quantities that correspond to the  linear limit). This problem is readily solved giving a pair of double-degenerate eigenvalues, $\tb_\pm=\pm\sqrt{\delta^2 - \kappa^2}$, each having an invariant subspace spanned by two $\PT$-conjugate  eigenvectors, $\tbA_\pm^{(1)}$ and $\tbA_\pm^{(2)}=\PT \tbA_\pm^{(1)}$:
\begin{equation}
\label{eq:A+}
\tbA_\pm^{(1)}=\left(\!\!\begin{array}{c}
\phantom{+}\kappa e^{i\varphi}\sin\alpha  \\%
-\kappa e^{i\vartheta}\cos\alpha \\%
 0 \\%
 i(\tb_\pm -\delta)
\end{array}\!\!\right),\,\,
\tbA_\pm^{(2)}=\left(\!\!\begin{array}{c}
-\kappa  e^{-i\vartheta}\cos\alpha  \\%
-\kappa e^{-i\varphi}\sin\alpha \\%
 i(\tb_\pm -  \delta) \\%
  0
\end{array}\!\!\right).
\end{equation}
These vectors are mutually orthogonal: $\langle \tbA_\pm^{(1)},\tbA_\pm^{(2)}\rangle=0$, where $\langle
\bA,{\bf B}\rangle=\bA^\dag{\bf B}$ defines the inner product. For some general properties of odd-$\PT$-symmetric Hamiltonians see~\cite{SmithMathur,BendKlev}.

The   odd $\PT$ symmetry does not exhaust all the symmetries of the system. In particular, the unitary transformation 
$\tilde{H}=S H_\delta S^{-1}$, where $S$ is the block matrix \cite{supplement}  $S=\mbox{diag} \left( e^{i(\vartheta-\varphi)\sigma_3/2},e^{-i(\varphi+\vartheta)\sigma_3/2}\right)$
results in an {\em even}-$\PT$-symmetric Hamiltonian $\tilde{H}$ with the same $\p$ operator and with conventional ``bosonic'' time-reversal $\K$: $[\tilde{H},\p\K]=0$. { Additionally, $H_\delta$  anti-commutes with the charge conjugation operator $\C=\left( \sigma_1\otimes e^{i\sigma_3(\varphi-\pi/2)}\right)\K$, this symmetry being responsible for the eigenvalues $\tb_\pm$ to emerge in opposite pairs which are either real  (unbroken phase, $|\kappa|<|\delta|$) or purely imaginary   (broken phase, $|\kappa|>|\delta|$)~\cite{referee}.}  

Another important property of the odd $\PT$ symmetry is the existence of integrals of motion which can be found even in the nonlinear case.
First, using that $\p H_\delta\p=H^\dagger_\delta$ and $\p F(\bA)\p= F^\dagger(\bA)$,
one straightforwardly verifies \cite{ZK} that $\cQ=\bA^\dagger\p\bA$ is constant: $d\cQ/dz=0$. This conservation law locks the  power imbalance in the waveguides: $\cQ = P_1 - P_2= \textrm{const}$, where $P_1=|A_1|^2 + |A_2|^2$ and $P_2= |A_3|^2 + |A_4|^2$.
Furthermore, system (\ref{main_dynamic}) has a Hamiltonian structure. Indeed, defining a real-valued Hamiltonian $
\cH = \bA^\dagger\p \left[H_\delta  - { F(\bA)}/{2}\right]\bA
$~\cite{supplement}, Eq.~(\ref{main_dynamic}) can be rewritten  as  $i\dot{A}_{1,2}=\partial\cH/\partial A_{1,2}^*$ and $i\dot{A}_{3,4}=-\partial\cH/\partial A_{3,4}^*$. 
Obviously, $\cH$ is another conserved quantity:  $d\cH/dz=0$. 

\paragraph{Coherent switch.} 
Now we turn to examples illustrating features of the introduced coupler. Returning to  the linear case,  we observe that the double-degeneracy of eigenstates is protected by the  odd $\PT$ symmetry, i.e.,  the degeneracy cannot be lifted by any change of the parameters preserving $\PT$ symmetry. Thus  manipulating  such a coupler, one simultaneously affects both  the modes with the same propagation constant. This suggests an idea  to  perform a switching between a {\em  superposition} of binary states, rather than between independent states 
as it happens with usual $\PT$-symmetric switches~\cite{switch}. We call this device a {\em coherent switch}.   Since the mentioned superposition can be characterized by a free parameter, such a system simulates a quantum switch for a superposition of states. 

\begin{figure}
	\centering
\includegraphics[width=\columnwidth]{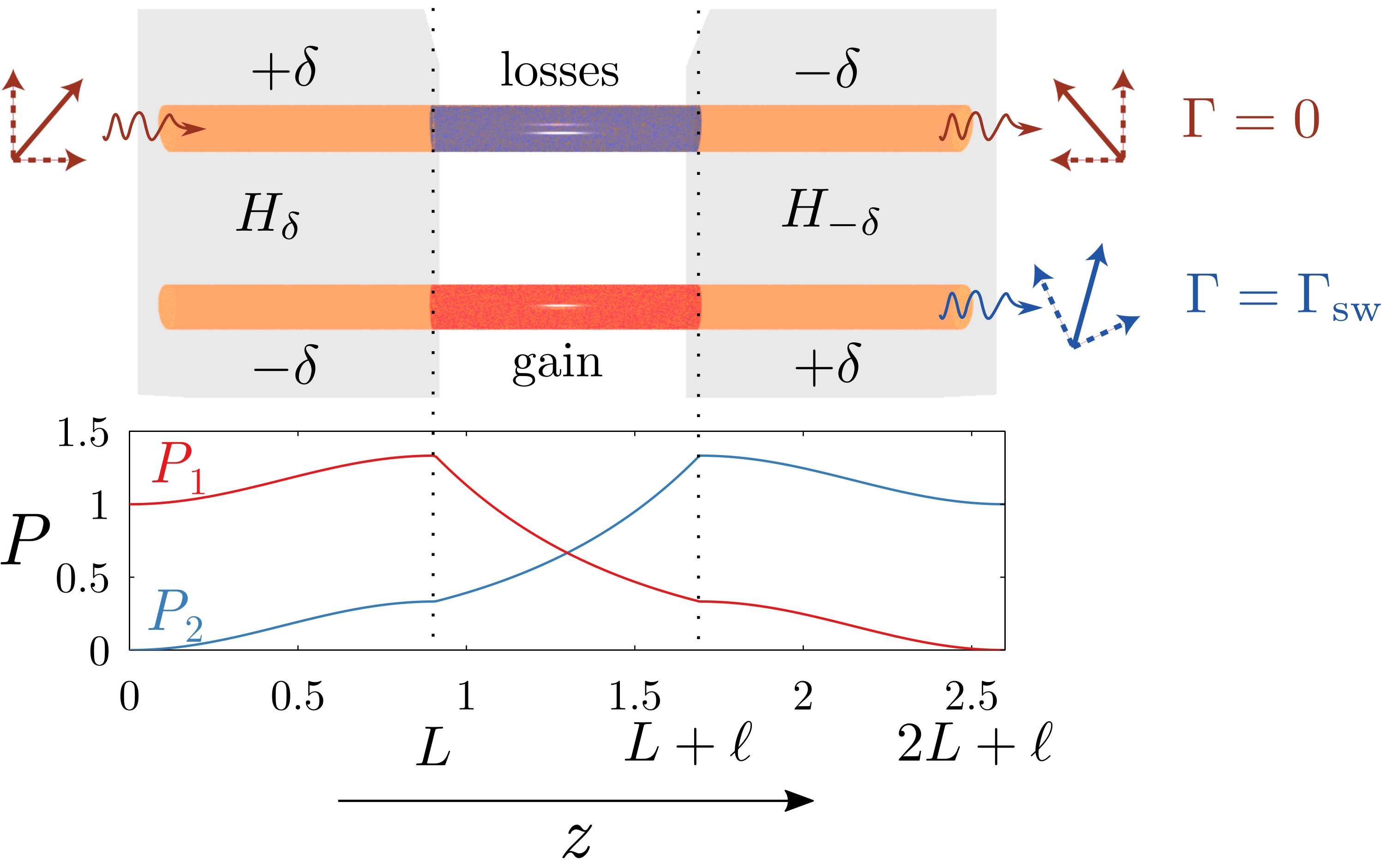}
	\caption{Upper panel shows schematically 
		the  coherent switch. Shadowed domains correspond to anti-$\PT$-symmetric media. The central empty part illustrates the uncoupled waveguides with gain and losses. The polarization vectors at the output indicate (schematically) 
		$\pi/2$-phase rotation of the non-switched signal (at $\Gamma=0$, red), and switching of the  superposition rotated by angle $-\alpha$ (at $\Gamma=\Gamma_{\rm sw}$, blue). Lower panel shows power distributions in the first $P_1$ (red line) and second $P_2$ (blue line) arms at $\Gamma_{\rm sw}$, obtained for $\delta=\delta_0=2$ and $\kappa=1$.    }
	\label{fig:switch}
\end{figure}

However, a solution for the coherent switch is not straightforward, because of the conservation of $\cQ$,
which means that an input signal, applied to only one waveguide, cannot be completely transfered to another  one. Since this conservation is due to the $\PT$ symmetry, the complete energy transfer between the arms is possible only if the symmetry is broken by an additional element 
%
at some propagation interval.  To this end, we explore the structure illustrated in 
Fig.~\ref{fig:switch}:  two couplers, with interchanged mismatches between the propagation constants, i.e., with $\delta\leftrightarrow-\delta$ in our notations, are connected by two decoupled waveguides. These auxiliary  waveguides have balanced losses $-\Gamma$ and gain $\Gamma$, and have a mismatch between the propagation constants, denoted by $\pm\delta_0$. The lengths of the couplers are equal and chosen as $L=\pi/(2\sqrt{\delta^2-\kappa^2})$ [we simplify the model letting $\vartheta=\varphi=0$]. The decoupled segment which disrupts the odd $\PT$ symmetry has the length $\ell=\pi/(2\delta_0)$. The propagation in the couplers is governed by $H_{\pm\delta}$, and can be expressed through the  evolution operators  
$U_{\pm\delta}(z,z+L) =  - i H_{\pm\delta}/\sqrt{\delta^2-\kappa^2}$~\cite{supplement}. The  evolution operator of the decoupled segment is diagonal: $U_0(z,z+\ell)=$diag$(ie^{-\Gamma \ell},ie^{-\Gamma \ell},-ie^{\Gamma \ell},-ie^{\Gamma \ell})$. Thus the output (at $z=2L+\ell$) and input (at $z=0$) fields are related by: 
\begin{equation}
\tbA_{\rm out}=U_{-\delta}(L+\ell, 2L+\ell)U_{0}(L,L+\ell)U_{\delta}(0,L)\tbA_{\rm in}. 
\end{equation}

The switch is controlled by the gain-and-loss coefficient $\Gamma$. Consider the situation when the input signal is   applied to the first waveguide and has the polarization $\tbA_{\rm in}=(\cos\chi,\sin\chi,0,0)^T$, i.e., $\tbA_{\rm in}$ is  parametrized by a free parameter $\chi$ (the red polarization vector at the input in Fig.~\ref{fig:switch}). If the waveguides in the central part are conservative, $\Gamma=0$, then the output signal is detected only at the first waveguide and arrives  $\pi/2$-phase-shifted: $\tbA_{\rm out}^{(0)}=i\tbA_{\rm in}$. If however $\Gamma=\Gamma_{\rm sw}=\ell^{-1}\ln(\delta/\kappa)$,  then the output signal has polarization  rotated by angle $-\alpha$  and  is detected only in the second waveguide: $\tbA_{\rm out}= (0,0,\cos(\chi-\alpha),\sin(\chi-\alpha))^T$ (blue polarization vectors in Fig.~\ref{fig:switch}). Importantly, $\chi$, i.e., the ratio  between the polarization components remains a free parameter. The power distributions in the waveguides in regime of   switching is shown in the lower panel of Fig.~\ref{fig:switch}. Inside the couplers,  both $P_{1,2}$ grow or  decay simultaneously. 
However, in the  central segment with disrupted odd $\PT$ symmetry the powers are adjusted in such a way that the complete energy transfer is observed   at the output. 
 
\paragraph{Nonlinear modes.}  
As the second example illustrating the unconventional features of our system, we consider peculiarities of modes guided  in a nonlinear coupler (\ref{main_dynamic})  with   odd-time $\PT$ symmetry. Stationary solutions are searched in the form $\bA = e^{-ibz} \ba $, where $b$ is a constant, and the amplitude vector $\ba$ solves the algebraic system  $b\ba = H_\delta\ba - F(\ba)\ba$. Since the nonlinearity is $\PT$ symmetric~\cite{ZK}, i.e., $[\PT,F(\ba)]=0$, the nonlinear modes with the same propagation constant appear in $\PT$-conjugate pairs: $\ba$ and $\PT\ba$.
Thus   
the nonlinearity does not lift the degeneracy, and   both $\PT$-conjugate modes are characterized by equal total powers $P=P_1+P_2=\ba^\dagger\ba$. The dependence $P(b)$ characterizes a family of modes; distinct families have different functional dependencies $P(b)$. Thus,  any result for a family $P(b)$ discussed below applies to the pair of  $\PT$-conjugate families. 

We start by analyzing how the nonlinearity affects linear modes, i.e., with the weakly nonlinear case. It is known \cite{general_p,SIAM}, that, in a system with an even $\PT$-symmetry without degeneracy of eigenstates, a linear eigenvalue bifurcates into a {\em single} family of nonlinear modes. But in a system with odd $\PT$ symmetry the situation can be more intricate, since   the eigenvalues are degenerate, and  one has to contemplate the effect of nonlinearity on a \textit{linear combination} of  independent eigenstates. The latter can be written as 
$\tbA_s =\sin(\nu) \tbA_s^{(1)} + \cos(\nu) e^{i\chi} \tbA_s^{(2)}$, where $\nu$ and $\chi$ are real parameters and $s$ stands for either ``$+$'' or ``$-$''.  Following~\cite{LZKK, ZK}, we look for a small-amplitude nonlinear mode in the form of  expansions $b_s = \tb_s  + \epsilon^2 \beta_s + \ldots$, and $\ba_s = \epsilon \tbA_s + \epsilon^3  {\bf {A}}_s^{(3)} + \ldots$, where $\epsilon\ll 1$ is a formal small parameter. From the $\epsilon^3$-order equation  we compute \cite{supplement}: $\beta_s= -\langle F(\tbA_s)\tbA_s^*, \tbA_s^{(j)}\rangle/\langle \tbA_s^*, \tbA_s^{(j)}\rangle$  which must be   satisfied for both $j=1,2$. Additionally, the coefficient $\beta_s$ is required to  be real.  These three requirements form the bifurcation conditions defining the parameters  $\nu$  and $\chi$ for which bifurcations of nonlinear modes are possible.

\begin{figure}
	\centering
	\includegraphics[width=1.0\columnwidth]{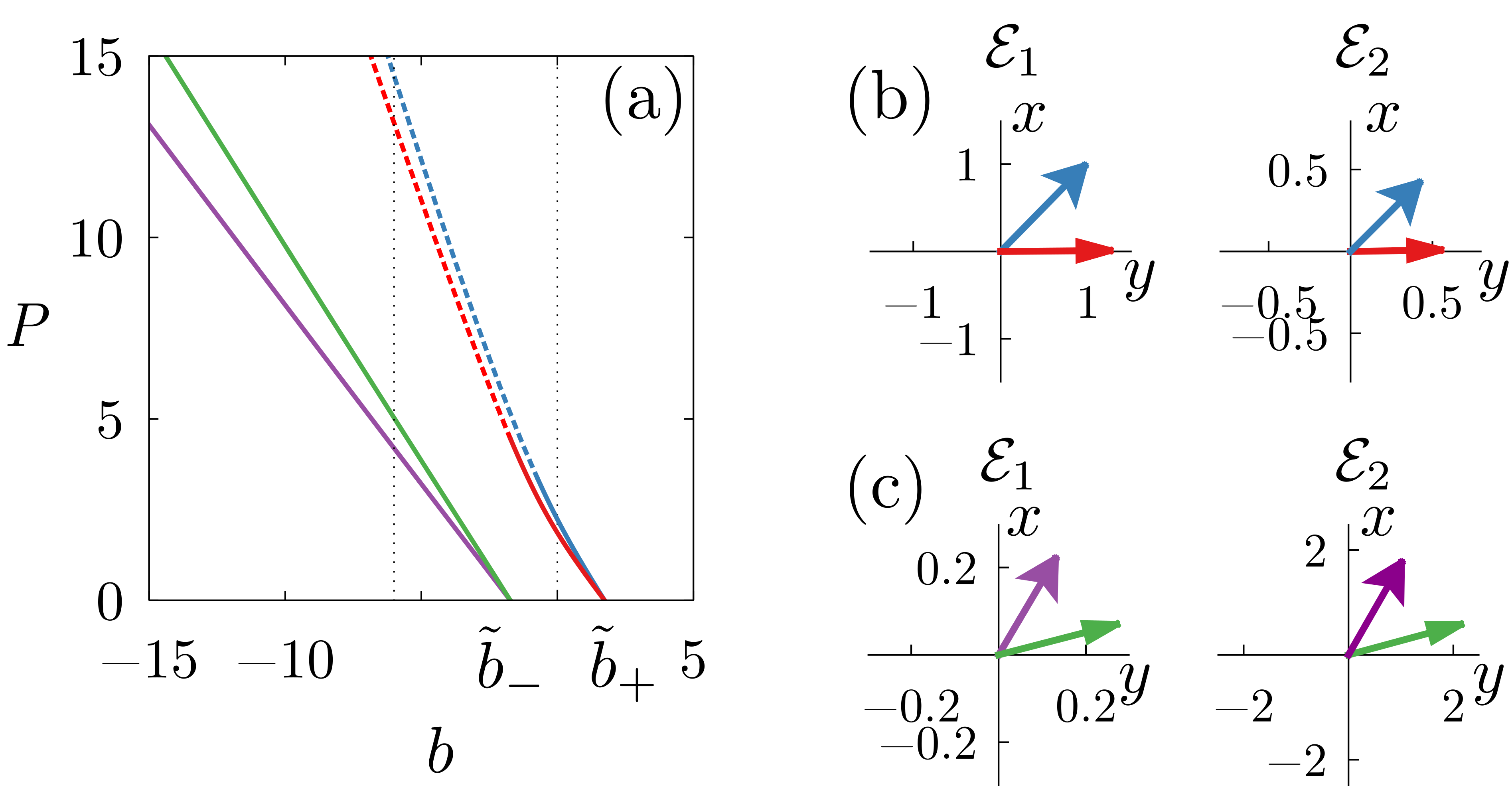}
	\caption{(a) Families  of solutions for  $\delta=2$, $\kappa=1$, $\alpha=\pi/6$, $\vartheta=\varphi=0$ visualized as dependencies $P$ {\it vs} $b$.  We show two families bifurcating from each eigenvalue $\tb_+$ and $\tb_-$. Stable and unstable modes are represented by solid and dotted segments, respectively. Vertical dotted lines indicate values $b=0$ and $b=-6$ analyzed in (b,c) and Fig.~\ref{fig:pol_kappa}. 	(b)Polarization vectors $\cE_{1,2}$ in each waveguide  for two stable nonlinear  modes bifurcating from $\tb_+$, at $b=0$.   Arrows corresponding to the same mode have the same color as the respective family (and the same arrow head). (c)  Polarization vectors $\cE_{1,2}$ for two stable modes bifurcating from 	 and $\tb_-$, at $b=-6$.   {The  lengths of arrows $\cE_{1,2}$  are equal to powers $P_{1,2}$ in each arm.}}
	\label{fig:two}
\end{figure}

Let us analyze the simple case of $\vartheta=\varphi=0$ and $\alpha\in(0, \pi/4)$ [$\alpha=0,\pi/4$ correspond to a trivial solution of parallel polarizations in the coupler arms]. Using computer algebra, one finds that the bifurcation conditions can be satisfied for two values of $\chi$. At $\chi=\pi/2$, nonlinear modes can bifurcate from the linear limit at  $\nu_0=\pi/4$. 
These modes, however, have been found unstable in the entire range of their existence. A 
more interesting case is realized when each $\tb_s$ gives birth to two stable families of nonlinear modes: these correspond  to $\chi=0$ and $\nu=\nu_s$ given by
$$
2\tan\nu_{s}= c_s\pm \sqrt{c_s^2+4}+ \sqrt{(c_s \pm \sqrt{c_s^2+4})^2 + 4},
$$ where
$
c_s = 8\delta \tb_s  (\delta- \tb_s)^2/[\kappa^4 \sin(4\alpha)]-2\tan(2\alpha).  
$
Using this analytical result, 
we performed numerical continuation of stable nonlinear modes from the small-amplitude limit to arbitrarily large amplitudes. Example of  the resulting diagram is shown in Fig.~\ref{fig:two}(a), where we present two power curves $P(b)$   bifurcating from each eigenvalues $\tb_+$ and $\tb_-$. 
Tracing the  dynamical stability of the modes along the power curves, we  have found that the families bifurcating from $\tb_-$ are stable in the entire explored range, while both families from  $\tb_+$ are stable for small powers and lose stability at large amplitudes.

To compute polarizations of the modes, we notice that the stable nonlinear modes $\ba$ bifurcating from the linear limit are $\p\K$ invariant, i.e., $\p\K\ba=\ba$. In our case this means that entries $a_{1,2}$ are purely real, and $a_{3,4}$ are purely imaginary. Thus one can construct real-valued  polarization vectors $\cE_1 = a_1 \be_1 + a_2 \be_2$ and $\cE_2 = -ia_3 \be_3 -i a_4\be_4$, where $\be_j$ are as defined above (see Fig.~\ref{fig:one}).
Polarization vectors for several stable nonlinear modes are shown in  Fig.~\ref{fig:two}(b,c). For each considered mode,  polarizations $\cE_1$ and $\cE_2$ are nearly, but not exactly, parallel in both waveguides,  and their direction    varies slightly  as the propagation constant changes. Thus the main impact of the growing total power $P$ is  the increase of  moduli of $\cE_1$ and $\cE_2$.  Fig.~\ref{fig:two}(b,c) also explains the main difference between nonlinear modes bifurcating from $\tb_+$ and $\tb_-$. In the former (latter) case most of the total power is concentrated in the first (second) waveguide, i.e., $P_1 > P_2$ and  $|\cE_1| > |\cE_2|$ ($P_2 > P_1$ and  $|\cE_2| > |\cE_1|$).

\begin{figure}
  	\centering
  	\includegraphics[width=1\columnwidth]{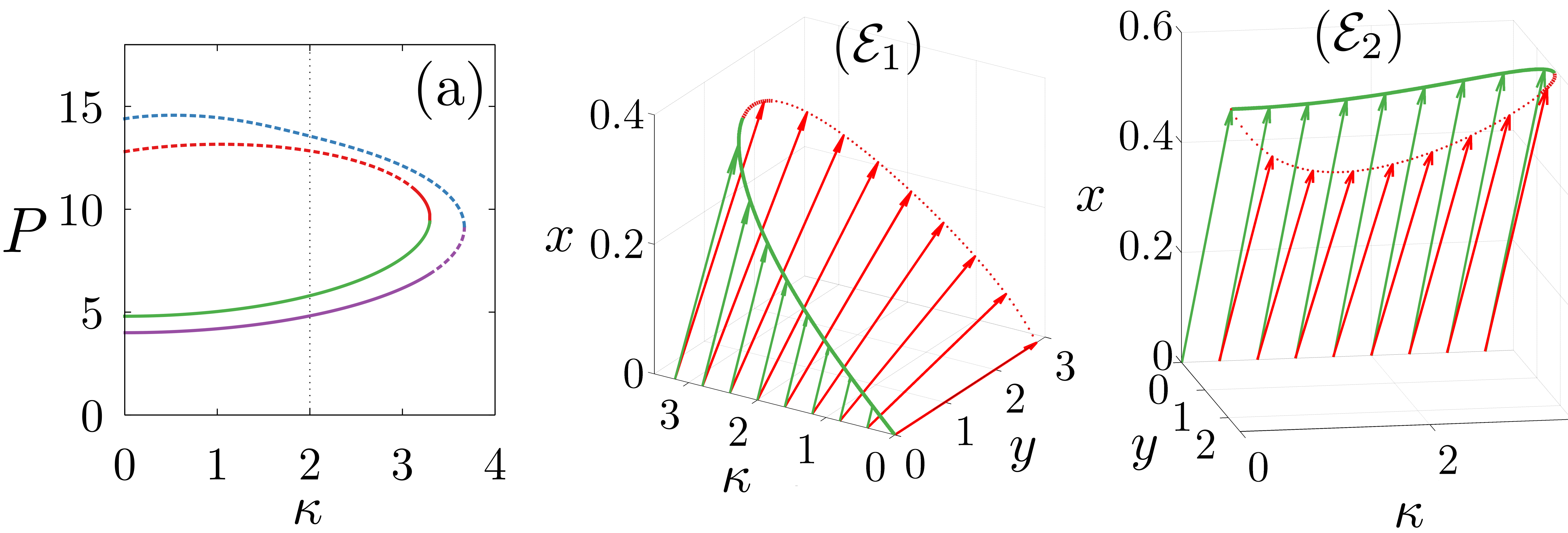}
  	\caption{(a) Branches of nonlinear modes for fixed $b=-6$, $\delta=2$, and changing $\kappa$. 
  	 Stable and unstable modes are represented by solid and dotted segments, respectively. Vertical dotted line indicates the $\PT$-symmetry-breaking threshold $\kappa=2$. 	 Panels ($\cE_{1,2}$)	 show  polarization vectors corresponding to two merging branches (red and green curves) from (a).
}
  	\label{fig:pol_kappa}
  \end{figure}

Figure~\ref{fig:pol_kappa}(a), where the dependencies $P$ {\it vs.} $\kappa$ are plotted for a fixed propagation constant,  illustrates the transformations of modes at the growing coupling strength $\kappa$. Four shown branches merge pairwise as $\kappa$ increases (each solution bifurcating from the positive eigenvalue $\tb_+$ merges with some solution  from $\tb_-$). Remarkably, the branches coalesce above the $\PT$-symmetry-breaking threshold $\kappa_\PT = \delta$ [which is equal to $2$ in Fig.~\ref{fig:two}(b)]. Moreover, solutions can be stable above the $\PT$-symmetry breaking point; in Fig.~\ref{fig:two}(a,b) stable modes are shown with solid lines. Polarization vectors of the nonlinear modes strongly depend on the coupling constant $\kappa$. This is illustrated in panels ($\cE_{1,2}$) of Fig.~\ref{fig:pol_kappa} where  the heads of vectors $\cE_{1,2}$ describe 3D curves in the $(\kappa,x, y)$ space. 
 
\textit{To conclude},  we  have introduced a $\PT$-symmetric optical coupler with 
odd-time reversal. The system features properties of an anti-$\PT$-symmetric medium in which two birefringent waveguides are embedded. As examples of  applications, we described a coherent switch which operates with a linear superposition of binary states with one free parameter. As the second example, we report on bifurcations of families of nonlinear modes. An unusual observation was  that each linear eigenstate gives raise to several distinct nonlinear modes,
some of which are stable. Although we dealt with an optical model, the way of architecture of $\PT$-symmetric systems is generic and can be implemented in other physical systems.  

\begin{acknowledgments}
	\textit{Acknowledgments}. The research of  D.A.Z. is supported by megaGrant No.~14.Y26.31.0015 of the Ministry of Education and Science of Russian Federation.
\end{acknowledgments}

\pagebreak
\widetext

\setcounter{equation}{0}
\setcounter{figure}{0}
\setcounter{table}{0}
\setcounter{page}{1}
\makeatletter
\renewcommand{\theequation}{S\arabic{equation}}
\renewcommand{\thefigure}{S\arabic{figure}}
\renewcommand{\bibnumfmt}[1]{[S#1]}
\renewcommand{\citenumfont}[1]{S#1}

\begin{center}
	\textbf{\large Supplemental Material for \textit{Odd-time reversal $\PT$~symmetry induced by anti-$\PT$-symmetric medium}}
\end{center}


\subsection{Some auxiliary expressions for formulation of the model}

In the expanded form, matrix $H_\delta$ reads
\begin{equation}
H_\delta = \left(  \begin{array}{cccc} %
\delta & 0 & i\kappa  e^{-i\vartheta}\cos\alpha & -i\kappa  e^{i\varphi}\sin\alpha\\
0 & \delta & i\kappa  e^{-i\varphi}\sin\alpha &  i\kappa  e^{i\vartheta}\cos\alpha\\
i\kappa  e^{i\vartheta}\cos\alpha & i\kappa  e^{i\varphi}\sin\alpha  & -\delta &0\\
-i\kappa  e^{-i\varphi}\sin\alpha & i\kappa  e^{-i\vartheta}\cos\alpha  &0 & -\delta
\end{array}
\right)
\end{equation}
Parity operator $\p$ (which is tantamount to  the Dirac $\gamma^0$ matrix) and   time-reversal operator $\T$ read:
\begin{equation}
\p = \gamma^0 = \left(  \begin{array}{cccc} %
1 & 0 & 0 & 0\\
0 & 1 & 0 & 0\\
0 & 0 & -1&0\\
0 & 0 &0 & -1
\end{array}
\right), \quad %
\T =  \left(  \begin{array}{cccc} %
0 & 1 & 0 & 0\\
-1 & 0 & 0 & 0\\
0 & 0 & 0&1\\
0 & 0 &-1 & 0
\end{array}
\right)\K,
\end{equation}
where $\K$ is the element-wise complex conjugation.

\bigskip

The real quaternion form of a matrix $C$ implies 
\begin{eqnarray}
C=c_0\sigma_0 +ic_1\sigma_1+ic_2\sigma_2+ic_3\sigma_3,
\end{eqnarray}
where  all $c_{0,...,3}$ are real and read
\begin{equation}
c_0 = \cos\alpha\, \cos\vartheta, \quad c_1 = -\sin\alpha\,\sin\varphi, \quad c_2 = -\sin\alpha\,\cos\varphi, \quad c_3 = -\cos\alpha\,\sin\vartheta.
\end{equation}

\bigskip

The explicit form of the matrix $S$ from the main text is
\begin{eqnarray}
S=\left(
\begin{array}{cccc}
e^{i(\vartheta-\varphi)/2} & 0 & 0 & 0 \\
0& e^{-i(\vartheta-\varphi)/2} &0 &0 \\
0 & 0& e^{-i(\vartheta+\varphi)/2} & 0 \\
0 & 0 & 0 & e^{i(\vartheta+\varphi)/2}
\end{array}\right)
\end{eqnarray}

\bigskip

The  Hamiltonian $
\cH = \bA^\dagger\p \left[H  - { F(\bA)}/{2}\right]\bA
$  can be expanded as 
\begin{eqnarray}
\cH = \delta(A_1A_1^*+A_2A_2^*+A_3A_3^*+A_4A_4^*)  \nonumber \hspace{9cm}\\[2mm]%
+  i\kappa\cos\alpha(A_1^* A_3e^{-i\vartheta} - A_1A_3^*e^{i\vartheta} + A_2^*A_4e^{i\vartheta} - A_2A_4^*e^{-i\vartheta}) \nonumber \hspace{5cm}\\[2mm]%
+i\kappa\sin\alpha(A_1A_4^*e^{-i\varphi} - A_1^*A_4e^{i\varphi} + A_2^*A_3e^{-i\varphi} - A_2A_3^*e^{i\varphi})\nonumber  \hspace{4cm}\\[2mm]
-\frac{1}{2}\bigl(|A_1|^4 + |A_2|^4 -|A_3|^4 - |A_4|^4\bigr)-\frac{2}{3}\bigl(|A_1|^2|A_2|^2 - |A_3|^2|A_4|^2\bigr).
\end{eqnarray}
The Hamiltonian equations read
\begin{eqnarray} 
\frac{\partial \cH}{\partial A_1^*} = i\dot{A}_1, \quad \frac{\partial \cH}{\partial A_2^*} = i\dot{A}_2, \quad \frac{\partial \cH}{\partial A_3^*} = -i\dot{A}_3, \quad \frac{\partial \cH}{\partial A_4^*} = -i\dot{A}_4.
\end{eqnarray}

\subsection{``Evolution'' matrix for the coherent switch}

Consider $i\dot{\bA} = H_{\pm\delta}\bA$, $|\delta|>|\kappa|>0$. Computing one more derivative and using that $H_{\pm\delta}^2 = (\delta^2-\kappa^2)I$, where $I$ is $4\times 4$ identity matrix, we obtain vector linear oscillator equation $\ddot{\bA} + (\delta^2-\kappa^2) \bA=0$. Its general solution is 
\begin{eqnarray}
 \bA(z) = \cos(\sqrt{\delta^2 - \kappa^2}z) \bA(0) +  (\delta^2 - \kappa^2)^{-1/2}\sin(\sqrt{\delta^2 - \kappa^2}\, z) \dot{\bA}(0).
\end{eqnarray}
  Therefore, the evolution operator $U_{\pm\delta}(z,0)$, defined 
   by $\bA(z)=U(0,z)\bA(0)$, has the form
\begin{equation}
U_{\pm\delta}(0,z) = \cos(\sqrt{\delta^2-\kappa^2}\, z)I - \frac{i\sin(\sqrt{\delta^2-\kappa^2}\, z)}{\sqrt{\delta^2-\kappa^2}}H_{\pm\delta}.
\end{equation}

\subsection{Details for the analysis of bifurcations of the nonlinear modes}

Handling the introduced asymptotic expansions in the standard way, i.e., collecting the terms with the same degree of $\epsilon$, it is easy to see that equations with $\epsilon$ and $\epsilon^2$ are satisfied automatically.  At $\epsilon^3$ one arrives at the equation 
\begin{equation}
\label{eq:3}
\beta_s \tbA_s + F(\tbA_s) \tbA_s = (H_\delta- \tb_s I)  \bA_s^{(3)},
\end{equation}
where $I$ is 4$\times$4 identity matrix.
Let $H_\delta^\dagger$ be an operator which is Hermitian conjugate to $H$. For $|\delta|>|\kappa|>0$ its eigenvalues are equal to those of $H$, i.e., amount to $\tb_+$ and $\tb_-$. Since for $\varphi=\vartheta=0$, $H_\delta$ and $H_\delta^\dagger$ are symmetric matrices (i.e., invariant under the transposition), any eigenvector of $H_\delta^\dagger$ is a linear combination of corresponding eigenvectors of $H_\delta$ taken with complex conjugation, i.e., amounts to ${\textbf D}_s = d_1 \tbA_s^{(1, *)} + d_2  \tbA_s^{(2, *)}$, where $d_1$ and $d_2$ are arbitrary constants, $|d_1|^2 + |d_2|^2\ne 0$, and the asterisk is the element-wise complex conjugation. In other words, $(H_\delta^\dag -\tb_s I){\textbf D}_s = 0$. We multiply both sides of (\ref{eq:3}) by ${\textbf D}_s$ (in the sense of the inner product $\langle\cdot, \cdot\rangle$) to obtain
\begin{equation}
\label{eq:aux}
\langle \beta_s \tbA_s + F(\tbA_s) \tbA_s,  d_1 \tbA_s^{(1,*)} + d_2  \tbA_s^{(2,*)} \rangle =0.
\end{equation}
Setting  in (\ref{eq:aux}) $d_1=0$, $d_2 =  1$  and $d_1= 1$, $d_2=0$, we conclude that $\beta_s$ has to satisfy two equations simultaneously (these equations are tantamount to those from the main text):
\begin{equation}
\beta_s= -\frac{\langle F(\tbA_s)\tbA_s^*, \tbA_s^{(1)}\rangle}{\langle \tbA_s^*, \tbA_s^{(1)}\rangle}, \quad \beta_s= -\frac{\langle F(\tbA_s)\tbA_s^*, \tbA_s^{(2)}\rangle}{\langle \tbA_s^*, \tbA_s^{(2)}\rangle},
\end{equation}
where additionally one has to require $\beta_s$ to be real ($\beta_s = \beta^*_s$) for the propagation constants of nonlinear modes to be real.


\begin{thebibliography}{99}
	
	\bibitem{BenderBoet}  C. M. Bender and S. Boettcher, 
	Real Spectra in Non-Hermitian Hamiltonians Having $\PT$-Symmetry.
	Phys. Rev. Lett. {\bf 80} 5243 (1998); C. M. Bender,
	Making sense of non-Hermitian Hamiltonians. 
	Rep. Prog. Phys. {\bf 70}, 947--1018 (2007).
	
	\bibitem{review}  V. V. Konotop, J. Yang, and D. A. Zezyulin,
	Nonlinear waves in $\PT$-symmetric systems. 
	Rev. Mod. Phys.  {\bf 88}, 035002 (2016).  
	
	\bibitem{Muga} A. Ruschhaupt, F. Delgado, and J. G. Muga,  
	Physical realization of $\PT$-symmetric potential scattering in a planar slab waveguide. 
	J. Phys A. {\bf 38}, L171 (2005).
	
	\bibitem{disc_opt}
	R. El-Ganainy, K. G. Makris, D. N. Christodoulides,  Z. H. Musslimani,  
	Theory of coupled optical PT-symmetric structures.
	Opt.  Lett.  \textbf{32},   2632 (2007).
	
	\bibitem{Christodoulides}
	Z. H. Musslimani,  K. G. Makris, R. El-Ganainy, D. N.  Christodoulides,  
	Optical Solitons in $\PT$ Periodic Potentials. 
	Phys. Rev. Lett. {\bf 100}, 30402 (2008);	
	K.G. Makris, R. El-Ganainy, D.N. Christodoulides, Z.H. Musslimani, Phys. Rev. Lett. {\bf 100}, 103904 (2008).
	
	\bibitem{BendManh}  C. M. Bender, P. N. Meisinger, and Q. Wang, Finite-dimensional $\PT$-symmetric Hamiltonians.
	J. Phys. A  {\bf 36}, 6791 (2003); A. Mostafazadeh, Exact $\PT$-symmetry is equivalent to Hermiticity, J. Phys. A {\bf 36}, 7081 (2003).
	
	
	
	\bibitem{general_p} K. Li and P. G. Kevrekidis,  $\PT$-symmetric oligomers: Analytical solutions, linear stability, and nonlinear dynamics. 
	Phys. Rev. E {\bf 83}, 066608 (2011); D. A. Zezyulin and V. V. Konotop, 
	Nonlinear Modes in Finite-Dimensional $\PT$-Symmetric Systems. 
	Phys. Rev. Lett. {\bf 108}, 213906 (2012); 
	K. Li, P. G. Kevrekidis, B. A. Malomed,  and U. G\"unther,  
	Nonlinear $\PT$-symmetric plaquettes. 
	J. Phys. A  {\bf 45}, 444021 (2012).
	
	
	
	\bibitem{ZK} D. A. Zezyulin and V. V. Konotop, Stationary modes and integrals of motion in nonlinear
	lattices with a $\PT$-symmetric linear part, J. Phys. A: Math. Theor. {\bf 46}, 415301 (2013).
	
	\bibitem{Messiah} A. Messiah, Quantum Mechanics, Volume II (John Wiley \& Sons, Inc. -- New York, 1966).
	
	\bibitem{SmithMathur} K. Jones-Smith and H. Mathur,  
	Non-Hermitian quantum Hamiltonians with PT symmetry. 
	Phys. Rev. A  {\bf 82}, 042101 (2010). 
	
	\bibitem{BendKlev} C. M. Bender and S. P.  Klevansky,   
	$\PT$-symmetric representations of fermionic algebras. 
	Phys. Rev. A  {\bf 84},  024102 (2011).
	
		\bibitem{switch} F. Nazari, M. Nazari, and M. K. Moravvej-Farshi,   
	A $\PT$ spatial optical switch based on $\PT$-symmetry. Opt. Lett. {\bf 36}, 4368 (2011); 
	A. Lupu, H. Benisty, and A. Degiron, 
	Switching using $\PT$-symmetry in plasmonic systems: positive role of the losses.
	Opt. Expr. {\bf 21}, 21651 (2013); 
	A. Lupu, H. Benisty, and A. Degiron,  
	Using optical $\PT$-symmetry for switching applications. 
	Photonics Nanostruct. Fundam. Appl. {\bf 12}, 305 (2014); 
	A. Lupu, V. V. Konotop, and H. Benisty,  Optimal $\PT$-symmetric switch features exceptional point, Sci. Rep. {\bf 7}, 13299 (2017).

	
	
	\bibitem{Ge} L. Ge and H. E. Tureci, 
	Antisymmetric $\PT$-photonic structures with balanced
	positive-negative-index materials. Phys. Rev. A 
	{\bf 88}, 053810 (2013). 
	
	\bibitem{Peng2016} P. Peng, W. Cao, C. Shen, W. Qu, J. Wen, L. Jiang and Y. Xiao,
	Anti-parity–time symmetry with flying atoms.
	Nat. Phys. {\bf 12}, 1139 (2016). 
	
	\bibitem{anti-PT} F. Yang, Y.-C. Liu, and L. You, Anti-$\PT$ symmetry in dissipatively coupled optical systems. Phys. Rev. A {\bf 96}, 053845 (2017).
	
		
	\bibitem{supplement} For the sake of convenience, in Supplemental Material we recall some definitions, present explicitly some of the matrices, and comment on details of the some algebra used in the main text. 
	
	\bibitem{Menyuk} C. R. Menyuk, Nonlinear pulse propagation in birefringent optical fibers, IEEE J. Quantum Electron. {\bf 23}, 174 (1987).
	
	
	\bibitem{SIAM} P. G. Kevrekidis,  D. E. Pelinovsky, and D. Y. Tyugin, SIAM
	J. Appl. Dyn. Syst. \textbf{12}, 1210 (2013).
	
	\bibitem{referee} {The existence of charge conjugation symmetry was noticed by the anonymous referee, who also pointed out that the Hamiltonian $H_\delta$ considered here belongs the class DIII of the classification introduced in S. Ryu, A. P. Schnyder, A. Furusaki, and A. W. W. Ludwig, Topological insulators and superconductors: tenfold way and dimensional hierarchy. New J. Phys. {\bf 12}, 065010 (2010).}
	
	\bibitem{LZKK} K. Li,  D. A. Zezyulin, V. V. Konotop, and P. G. Kevrekidis, Parity-time-symmetric optical coupler with birefringent arms. Phys. Rev. A {\bf 87}, 033812 (2013).

		
\end{thebibliography}
\end{document}